\documentclass[onecolumn, aps, prd, amsmath, amssymb, 12pt]{revtex4-2}

\usepackage{graphicx}
\usepackage{dcolumn}
\usepackage{bm}
\usepackage{amsfonts}
\usepackage{amsmath}
\usepackage{amssymb}
\usepackage{graphicx,color}
\usepackage{xcolor}
\usepackage{float}
\usepackage{hyperref}
\usepackage{subfigure}
\usepackage{mathtools,slashed}
\usepackage{blindtext}
\usepackage{longtable}
\usepackage{dcolumn}
\usepackage{bm}
\usepackage{braket}
\usepackage{appendix}
\usepackage{multirow}
\usepackage{color}
\usepackage{soul}

\usepackage{lipsum}
\usepackage{lmodern}	
\usepackage[T1]{fontenc}
\usepackage[utf8]{inputenc}

\usepackage{hyperref}
\hypersetup{
    colorlinks=true,
    linkcolor=blue,
    filecolor=red,      
    urlcolor=blue,
    citecolor=red
}

\begin{document}
\title{LQG-modified dispersion relations and the problem of cosmic photons threshold anomalies}

\author{P. A. L. Mourão}
\email{pedrolimamourao@gmail.com}
\affiliation{Centro Brasileiro de Pesquisas F\'{\i}sicas, Rua Dr. Xavier Sigaud 150, Urca, CEP 22290-180, Rio de Janeiro, RJ, Brazil}

\author{G. L. L. W. Levy}
\email{guslevy9@hotmail.com}
\affiliation{Centro Brasileiro de Pesquisas F\'{\i}sicas, Rua Dr. Xavier Sigaud 150, Urca, CEP 22290-180, Rio de Janeiro, RJ, Brazil}

\author{J. A. Helayël-Neto} 
\email{josehelayel@gmail.com}
\affiliation{Centro Brasileiro de Pesquisas F\'{\i}sicas, Rua Dr. Xavier Sigaud 150, Urca, CEP 22290-180, Rio de Janeiro, RJ, Brazil}

\begin{abstract}
\hspace{0.05cm}Our present contribution sets out to investigate how combined effects of Lorentz invariance violation (LIV) and loop quantum gravity (LQG)-modified photon dispersion relations affect the threshold anomaly of cosmic photons. The point of departure is the post-Maxwellian version of electromagnetism induced by LQG effects. We then consider the problem of gamma-ray attenuation by the extragalactic background light (EBL) and the cosmic microwave background radiation (CMB) dominated by the Breit-Wheeler Effect. By following this path, we aim at the establishment of a new bridge between LIV and astrophysical phenomena in the framework of LQG, providing a robust explanation for recent ultra high energy (UHE) astrophysical observations.

\vspace{0.4cm}

keywords: loop quantum gravity, Lorentz Invariance Violation, Threshold Anomalies, Modified Dispersion Relations, UHE Astrophysical Observation.
\end{abstract}

\maketitle 

\section{Introduction}
\label{sec1}
Lorentz invariance is one of the foundation stones of (relativistic) quantum field theories, elementary particle physics and, more recently, there have even appeared new condensed matter systems that exhibit an emergent Lorentz invariance~\cite{kostelecky1998, kostelecky2011} . The dynamics of the Standard Model of Fundamental Interactions and Particle Physics (SM) does not give room to the breaking of Lorentz invariance. In his seminal 1951 Nature's paper, ``Is there an Aether?"~\cite{dirac}, Paul Dirac contemplates a physical scenario where Lorentz-invariance violating effects may become relevant. He argued that \textit{``we are rather forced to have an aether"}. Ever since, the discussion on Lorentz-invariance violation (LIV) has been strengthened in the physical literature, with a remarkably intense activity over the past three decades. This is motivated by the fact that Lorentz-violating effects are predicted by a number of contemporary models built up in the search for Physics Beyond the Standard Model. Among such models we can mention string theory~\cite{kostelecky1989, ellis1999, ellis2000, ellis2004}, doubly special relativity~\cite{camelia2002, camelia2005, glikman2005}, supersymmetry and supergravity~\cite{wess1974, martin1998, helayel2016, helayel2019, helayel2021} and some approaches to quantum gravity such as loop quantum gravity~\cite{pullin1999, alfaro2000, alfaro2002-1, alfaro2002-2, levy2024, melo2024, mourao2025}, to name a few. Up to now, we know that LIV has not yet been experimentally confirmed, despite all the high-precision experiments currently carried out~\cite{matingli1}. However, we keep in mind the importance of considering the possibility of its detection as the searches become more refined. Since terrestrial accelerators have not yet managed to detect the tiny LIV effects, astrophysical environments where the production of ultra high energy phenomena occur turn out to be an excellent field of investigation to inspect LIV~\cite{matingli2,jacob1,Li2,Li4,Li3}.

Ground-based detectors such as the Large High Altitude Air Shower Observatory (LHAASO)~\cite{lhaaso} in China and the High-Altitude Water Cherenkov Observatory (HAWK)~\cite{hawk} in Mexico have detected very high energy (VHE) and ultra high energy (UHE) photons. These detections challenge the limits of special relativity (SR), which impose an opacity window for a considerable range of these photons. This phenomenon occurs due to pair production by the kinematics of the Breit-Wheeler effect (a phenomenon investigated in the seminal 1934 article in Ref.~\cite{breit1934}), in which gamma photons, representing the most energetic range of the electromagnetic spectrum, are attenuated by photons from the background lights of the Universe, such as extragalactic background light and cosmic microwave background, generating an electron-positron pair $(\gamma \gamma \rightarrow e^+e^-)$. This phenomenon prevents the Universe from being transparent to the free propagation of these photons, especially at the PeV scale. However, the laboratories mentioned, among others, have detected photons in these energy ranges, although many cases are inconsistent with the thresholds established by special relativity.

The aim of this work is to derive a modified dispersion relation (MDR) from an effective hamiltonian in the semiclassical regime of the electromagnetic sector of loop quantum gravity (LQG). Once this MDR is established, we can deduce a cubic threshold equation that, at the zero discriminant, establishes the critical thresholds for EBL and CMB. The analysis of the discriminants of this equation makes it possible to establish the relationship between the effective parameter of the Lorentz violation of the LQG and the critical thresholds of the cosmic backgrounds. These relationships reveal phenomena such as the opacity window, discussed earlier, and the anomalous transparency for VHE and UHE photons.

In 2021, LHAASO detected the highest-energy photon ever recorded, at 1.42 PeV, from the LHAASO J2032+4102 source, in Cygnus Cocoon, a region of our galaxy with intense star formation~\cite{LHAASO2021}. The highlight of our model is explaining how the record breaking energy photon was detected by LHAASO and not strongly suppressed by the CMB as predicted by SR. Furthermore, it is important to emphasize that the consequence of strong anomalous transparency for UHE photons, caused by the spacetime deformation by the Lorentz invariance violation parameter of the LQG, does not establish a limit for the mean free path $\lambda$ of these photons, expanding the gamma-ray horizon of special relativity. In this way, we wish to demonstrate that the threshold anomalies induced by the LQG MDR provide a consistent explanation for the photon detected by LHAASO.

Our paper is organized according to the following scheme. In Section II, we deduce a LQG Modified Dispersion Relation from an effective hamiltonian in the semiclassical regime. In Section III, we introduce all the necessary Theoretical Foundations for the development of this research, defining the concept of threshold anomalies induced by LIV and then deducing a cubic threshold equation of the LQG, performing a qualitative analysis of the discriminants that allowed us to determine an expression for the critical point of the LIV factor of background radiations. In Section IV, we begin with the Results and Discussions, this time performing a quantitative analysis of the discriminants, followed by an cutting-edge observational connection. Finally, in Section V, we present the Conclusions regarding the development of this work.
 
\section{LQG-modified dispersion relation}
\label{sec4}
As discussed in the introduction, in the present paper, we are motivated to place LQG effects in the study of VHE and UHE cosmic photons, more specifically, the issue of cosmic-photon threshold anomalies. To explain the anomalous transparency mentioned in the introduction, we pick out a particular electrodynamic model, namely, a set of modified Maxwell-like equations with corrections induced by LQG, which yield a set of field equations that describe the propagation of electromagnetic radiation at the Planck scale. Through these equations, we will find a dispersion relation modified by the LQG, which will be simplified until we explicitly reveal the LIV factor of the adopted quantum gravity model. More precisely, we start off from the hamiltonian of the electromagnetic field extended by the presence of parameters introduced by LQG. This hamiltonian is cast below, and the whole procedure for its construction may be found in the detailed work by Alfaro et al. in Ref.~\cite{alfaro2000, alfaro2002-1}:
\begin{align}\label{ham}
H_{LQG} &= \frac{1}{Q^2} \int d^3x \Biggl \{\left[1+\theta_7 \left (\frac{\ell_p}{\mathcal{L}}\right)^{2+2\Upsilon}\right]\frac{1}{2} (\underline{\vec{B}}^2 + \underline{\vec{E}}^2)
+ \theta_3 \ell_p^2 (\underline{B}^a \nabla^2 \underline{B}_a + \underline{E}^a \nabla ^2 \underline{E}_a)\nonumber \\
& + \theta_2 \ell_p^2 \underline{E}^a \partial_a \partial_b \underline{E}^b +  \theta_8 \ell_p[\underline{\vec{B}}\cdot  (\nabla \times \underline{\vec{B}})+\underline{\vec{E}}\cdot  (\nabla \times \underline{\vec{E}})]  + \theta_4 \mathcal{L}^2 \ell_p^2 \left (\frac{\mathcal{L}}{\ell_p}\right)^{2\Upsilon} (\underline{\vec{B}}^2)^2 + \cdots  \Biggl\}.
\end{align}

Since loop quantum gravity requires extremely high energies to access the predicted effects, the hamiltonian above was extended to the semiclassical regime of gravity~\cite{thiemann2001, sahlmann2002}. Having done this, we observe effective classical electromagnetic fields, but the structure of spacetime remains quantum~\cite{gambini1999}. In this case, the LQG has edges, areas, and volumes related to $\ell_p$, $\ell_p^2$, and $\ell_p^3$ ~\cite{rovelli1995, ashtekar1997}, respectively, accompanied by effective corrections in the semiclassical regime. This allows the theory to be connected to observation. These corrections are parameterized by the terms $\theta_i$, with $i=2,3,4,7$. They are dimensionless and non-degenerate~\cite{alfaro2002-1, matingli1}.

The factor $Q^2$ is the electromagnetic coupling constant, and $\ell_p \approx 1.6 \times 10^{-35} \, m$ is the Planck scale, where quantum mechanics and general relativity unify in the LQG. Another important factor is the characteristic scale $\mathcal{L}$, which represents the size of the quantum mesh of the effective LQG discrete space and lies in the region $\ell_p \ll \mathcal{L} \lesssim \lambda$. The value $\lambda$ is the wavelength of the photon from the de Broglie relation. The term $\Upsilon$ is a hamiltonian correction exponent related to the energy scale that arises from the effective dynamics of the LQG; $a$ and $b$ are the indices of the tensors to which they are related.

Through the effective hamiltonian~\eqref{ham}, we find the sourceless Àmpere-Maxwell equation and the Faraday-Lenz equation, derived in ~\cite{alfaro2002-1}. They are represented, respectively, by
\begin{eqnarray}\label{ma1}
A_{\gamma}\left(\bm{\nabla} \times \vec{B} \right) &-&  \frac{\partial 	\vec{E}}{\partial t} + 2 \theta_3 \ell_p^2 \bm{\nabla}^2 \left( \bm{\nabla} 	\times \vec{B} \right) - 2\theta_8 \ell_p \bm{\nabla}^2 \vec{B}\nonumber \\
&+& 4 \theta_4 \ell_p^2 \mathcal{L}^2 \left(\frac{\mathcal{L}}{\ell_p}\right)^{2\Upsilon} \bm{\nabla} \times \left( \vec{B}^2 \vec{B}\right)=0,
\end{eqnarray}
\begin{eqnarray}\label{ma2}
A_{\gamma}\left( \bm{\nabla} \times \vec{E} \right) + \frac{\partial \vec{B}}{\partial t} + 2 \theta_3 \ell_p^2 \bm{\nabla}^2 \left( \bm{\nabla} \times \vec{E} \right)
 - 2\theta_8 \ell_p \bm{\nabla}^2 \vec{E}=0,
\end{eqnarray}
with
\begin{equation}\label{A}
A_{\gamma} = 1 + \theta_7 \left(\frac{\ell_p}{\mathcal{L}}\right)^{2+2\Upsilon}.
\end{equation} 
The solution of the equations $\eqref{ma1}$ and $\eqref{ma2}$ is complemented by the condition 
\begin{eqnarray}\label{zero}
\bm{\nabla} \cdot \vec{E}= \bm{\nabla} \cdot \vec{B}=0.
\end{eqnarray}

To solve the source-free Maxwell equations corrected by the LQG in the semiclassical regime, we will disregard the nonlinear term in the Àmpere-Maxwell equation and use plane wave solutions, given by
\begin{eqnarray}\label{op}
\vec{E}= \vec{E}_0 e^{i(\vec{k}\vec{x}- \omega t)}, \hspace{0.2cm} \vec{B}= \vec{B}_0 e^{i(\vec{k}\vec{x}- \omega t)}, \hspace{0.2cm} k:=|\vec{k}|.
\end{eqnarray}
Applying them to equations $\eqref{ma1}$ and $\eqref{ma2}$, we obtain the following relationships:
\begin{eqnarray}\label{e}
\left(\vec{k} \times \vec{E}_0 \right)\left[ A_{\gamma} - 2\theta_3 (\ell_p \vec{k})^2 \right] - 2i\theta_8 \ell_p \vec{k}^2 \vec{E}_0 - \omega \vec{B}_0=0,
\end{eqnarray}
\begin{eqnarray}
\label{b}
\left(\vec{k} \times \vec{B}_0 \right)\left[ A_{\gamma} - 2 \theta_3 (\ell_p \vec{k})^2 \right]- 2i\theta_8 \ell_p \vec{k}^2 \vec{B}_0 + \omega \vec{E}_0=0.
\end{eqnarray}
The solution to these equations leads us to a loop quantum gravity-modified dispersion relation, given by
\begin{equation}\label{rdm2}
\omega = k \left[A_{\gamma} - 2\theta_3 (k\ell_p)^2 \pm 2\theta_8 (k\ell_p)\right].
\end{equation}

We will simplify this MDR by first disregarding the term $\theta_3$, since it is multiplied by the quadratic Planck scale $\ell_p$. As this parameter is usually small, it undergoes a process of quantum dissipation, so that
\begin{equation}\label{rdm2}
\omega = k \left[A_{\gamma} \pm 2\theta_8 (k\ell_p)\right].
\end{equation}
The symbol $\pm$ associated with $\theta_8$ refers to the helicity of the photon, which generates different polarization modes for the electromagnetic wave, leading to a delay or advance in the propagation speed. Therefore, this parameter leads to birefringence effects. The best constraint in the literature is given by $\theta_8 \leq 10^{-16}$ \cite{macione1}. Since $k\ell_p \ll 1$, we can consider $\theta_8 k\ell_p \approx 0$. In this way, the MDR becomes 
\begin{equation}\label{rdm2}
\omega = k \left[1+\theta_7 \left(\frac{\ell_p}{\mathcal{L}}\right)^{2+2\Upsilon}\right].
\end{equation}
Now, we must linearize this relationship in the Planck scale term $\ell_p$, keeping it at the first order of the correction. To do this, we will set $\Upsilon=-1/2$. This is a necessary restriction to preserve the best possible access of the LQG to the effective semiclassical regime and experimental measurements
\begin{equation}\label{rdm2}
\omega = k \left[1+\theta_7 \left(\frac{\ell_p}{\mathcal{L}}\right)\right].
\end{equation}
The characteristic scale has its maximum at $\mathcal{L} \lesssim \lambda$. This value is precisely the wavelength of the photon in the de Broglie relation. Therefore, choosing $\mathcal{L} \approx 1/k$ preserves the factor $\ell_p k$ and addresses the granularity of spacetime in the propagating wave regime~\cite{amelino2013}. Thus, the dispersion relation becomes
\begin{equation}\label{rdm2}
\omega = k \left[1+\theta_7\left( \ell_p k \right)\right]
\end{equation}
Again, we consider only the linear order of Planck Scale. Thus, the dispersion relation modified by the LQG can be written as
\begin{equation}\label{rdl}
\omega^2 = k^2 + 2\theta_7 \ell_p k^3.
\end{equation}
To simplify the notation, we will write
\begin{equation}\label{lq}
\xi_{LQG} = - 2\theta_7 \ell_p.
\end{equation}
However, in the following Section, this non-arbitrary parameter inherited from the modified dispersion relation will be denoted simply $\xi$, in order to deduce a cubic threshold equation that is associated with a LIV factor. We will proceed with the necessary theoretical foundation to address the problems mentioned in the introduction of this work. 

\section{Theoretical Basis}
In this section, we will lay the necessary theoretical foundations for our entire phenomenological analysis. We will begin by differentiating the Breit-Wheeler effect from the perspective of special relativity and a dispersion relation modified by a LIV parameter. Starting from the MDR in $\eqref{rdl}$, where we replace $\xi_{LQG}$ with $\xi$, we will deduce a cubic threshold equation that, at the zero discriminant, provides us with a critical point related to cosmic backgrounds. Finally, we will perform a qualitative analysis of the discriminants, establishing how the LIV parameter of the LQG and the critical points of EBL and CMB radiation are related. Lastly, we will have a definition of opacity window and anomalous transparency for cosmic photons.

\subsection{Threshold Anomalies}
In theories like special relativity, pair production can occur in particle scattering with the usual scattering relation of the theory, such as $\omega^2 = k^2$. In this context, there is a lower energy threshold, below which processes are kinematically forbidden.

An example is the kinematics of the Breit-Wheeler effect~\cite{breit1934}, explained in the introduction of this work. The attenuation of gamma photons in this process generates a spectral cutoff that can be seen through the analysis of data from detectors of astrophysical sources. Besides LHAASO and HAWK, we have the Major Atmospheric Gamma Imaging Cherenkov Telescopes (MAGIC)~\cite{magic}, the High Energy Stereoscopic System (HESS)~\cite{hess}, among others. In addition to the ground-based ones, there is the satellite The Fermi Gamma-ray Space Telescope (FermiLAT)~\cite{fermi}, inaugurated in 2008 by NASA. We cannot fail to mention The Cherenkov Telescope Array (CTA)~\cite{cta}, a new observatory, the most sensitive among all, with more than one hundred planned detectors, which is under development and will open up a great opportunity in the study of gamma rays, improving the current detection capacity.

However, in kinematics with modified dispersion relations, where Planck-scaling suppressed terms associated with cubic momentum are introduced, the pair production process can behave somewhat differently from the kinematics of special relativity. As predicted by LIV-modified dispersion relations in the works of Mantingly et al.~\cite{matingli2} and Jacobson et al.~\cite{jacob1}, since the lower energy threshold can shift, in addition to the possibility of generating an upper energy threshold, or even cases where the process is completely suppressed. This result in the phenomenon of anomalous transparence, which we will investigate later. It is important to mention that the conservation laws remain valid, but the energy-momentum relation is altered.

The most fascinating question when using these dispersion relations with terms associated with cubic momentum is the simultaneous possibility of a lower and upper energy threshold, first studied by Kluniak~\cite{kluzniak}, generating an energy window that allows pair production. This happens because, in the process, using the conservation laws and the modified dispersion relation, we find a higher-order polynomial equation, whose solution allows us to find several real and positive roots. This is what makes the aforementioned phenomena possible, very different from processes that purely use special relativity.

A clear example of this modified dispersion relation is Eq.~\eqref{rdl}, whose modifications in the cubic term by the LQG in the semiclassical regime can be associated with effective violations of Lorentz invariance. From now on, we will use this specific MDR to deduce the polynomial equation we mentioned earlier, which may provide us with threshold anomalies induced by LIV. In this process, we will assume that the electron and positron will be described purely by special relativity. This is sufficient for the analysis that follows, based on Refs.~\cite{macione2007, ellis2008, ellis2009}.

\subsection{Cubic Threshold Equation and the Critical Point}
As discussed in the introduction to this Section, we will use the kinematic process of the Breit-Wheeler effect, schematically defined by
\begin{equation}
\gamma(\omega, k) \gamma(\epsilon_b) \rightarrow e^+ e^- .
\end{equation}
Where $\omega$ and $k$ are the MDR and momentum associated with the gamma photon, respectively, and $\epsilon_b$ is the photon energy of the background radiation. In the LQG-modified dispersion relation, starting from the Eq.~\eqref{ham}, we will use only $\xi$, as mentioned at the end of Sec~\ref{sec4}. This MDR can be expanded, to first order, as
\begin{equation}\label{xi}
\omega \approx k-\frac{\xi}{2}k^2.
\end{equation}
Since the photons from the background lights have much lower energies compared to gamma radiation, we will disregard the effects of LIV and write their energies only as $\epsilon_b$, because we need very high energies to access these effects.

The energy and momentum conservation relations are given by:
\begin{equation}
\omega + \epsilon_b = 2 \sqrt{p^2 + m_e^2},\label{cons}
\end{equation}
\begin{equation}\label{consp}
k- \epsilon_b = 2p,
\end{equation}
where the energy $E_{-} = E_{+} = E$ and the momentum $p_{-} = p_{+} = p$ of the electron and positron are considered equal. In the $\eqref{consp}$ relation, it is convenient that we do
\begin{equation}\label{consp1}
p = \frac{k- \epsilon_b}{2}.
\end{equation}
Now, we can rewrite the conservation of energy only in terms of momentum $k$, substituting momentum $\eqref{consp1}$ and the expansion $\eqref{xi}$ into the conservation relation $\eqref{cons}$, so that
\begin{eqnarray}\label{cons2}
k- \frac{\xi}{2}k^2 + \epsilon_b = 2\sqrt{\left(\frac{k - \epsilon_b}{2}\right)^2 + m_e^2}.
\end{eqnarray}
Raising both sides to the square, we have
\begin{eqnarray}\label{cons3}
\left(k- \frac{\xi}{2}k^2 + \epsilon_b \right)^2 = 4\left(\frac{k - \epsilon_b}{2}\right)^2 + 4m_e^2.
\end{eqnarray}
In the regime where $k \gg \epsilon_b $, the equation $\eqref{cons3}$ becomes
\begin{equation}\label{ec}
\xi k^3-4\epsilon_b k +4m_e^2=0,
\end{equation}
which is the cubic equation for threshold anomalies. Note that in the limit $\xi \rightarrow 0$, the cubic equation returns to the lower energy threshold of SR, given by
\begin{equation}\label{sr}
k_{th}^{SR} \geq \frac{m_e^2}{\epsilon_b},
\end{equation}
where only photons with energy greater than or equal to this threshold generate electron-positron pairs, culminating in the so-called opacity of the Universe. Photons with lower energy propagate freely.

Dividing the general cubic equation by $\xi$, we have
\begin{equation}
k^3 - \frac{4\epsilon_b }{\xi}k + \frac{4m_e^2}{\xi}=0.
\end{equation}
This equation is in the form
\begin{equation}
k^3 + qk + r=0.
\end{equation}
The discriminant of this equation
\begin{equation}
\Delta = - 4 q^3 - 27 r^2.
\end{equation}
Thus, we have:
\[
q = -\frac{4\epsilon_b}{\xi},
\]
\[ 
r = \frac{4m_e^2}{\xi}.
\]
Through these coefficients, the discriminant can be written as:
\begin{equation}\label{de}
\Delta = \frac{1}{\xi^3} \left(256 \epsilon_b^3 - 432 m_e^4 \xi\right).
\end{equation}
The critical point is given when the discriminant $\Delta=0$, which means a tangent line at the minimum inflection point of the function $f(k)$, with $k>0$. The physically important value occurs when
\begin{equation}\label{d0}
\Delta = 256 \epsilon_b^3 - 432 m_e^4 \xi_c = 0,
\end{equation}
where the critical point of the threshold condition is
\begin{equation}\label{lc}
\xi_c =	\frac{16}{27}\frac{\epsilon_b^3}{m_e^4}.
\end{equation}

We can also find this same result by differentiating the cubic function at zero, that is, $f'(k)=0$. Something similar was done by H. Li and Ma in Ref.~\cite{Li4}, who established a function $\xi(k)$ to determine the critical moment $k_c$ and, consequently, the critical point $\xi_c$. The distinguishing feature of this work was maintaining the cubic threshold equation as $f(k)=0$ and using a discriminant equal to zero to find the critical point. Therefore, we will also proceed with a qualitative analysis of the positive and negative discriminants.

\subsection{Qualitative Analysis of Discriminants}
In the cubic threshold function, we know that the cubic term $(\xi k^3)$ deforms spacetime by dominating LIV effects in the gamma energy range, the linear term $(4 \epsilon_b k)$ is the contribution representing the collision of high-energy photons with EBL and CMB background photons, while the electron mass is a constant term. The way the cubic and linear terms compete with each other determines the physical analysis of the function~\cite{arfken}.

Now, let's analyze the behavior of the discriminants of the cubic threshold function, starting for $k \in \mathbb{R}$:
\begin{itemize}
\item $\Delta>0$ - three real roots;
\item $\Delta=0$ - three real roots, with two repeated;
\item $\Delta<0$ - one real root and two complex conjugates.
\end{itemize}

Evaluating the same discriminants, but for $k>0$, which is the physically relevant regime, we have:
\begin{itemize}
\item $\Delta>0$ - two real and positive roots;
\item $\Delta=0$ - the curve touches a minimum;
\item $\Delta<0$ - no positive roots.
\end{itemize}

To understand the relationship between $\xi$ and $\xi_c$ with the sign of the determinant, we must add and subtract the numerator of $\eqref{de}$ by $432 m_e^4 \xi_c - 432 m_e^4 \xi$, so that
\begin{equation}\label{di}
\Delta = \frac{432 m_e^4}{\xi^3}\left(\xi_c - \xi \right).
\end{equation}
Now, with the expression $\eqref{di}$, we can directly interpret the relationship between $\xi$ and $\xi_c$ for $k>0$. Furthermore, we will qualitatively analyze the behavior of the cubic threshold function $\eqref{ec}$ using Fig. 1.
\begin{itemize}
\item \textbf{Case I (blue)}: $\xi < \xi_c $. \textbf{The discriminant is positive}. This region represents: opacity window;
\item \textbf{Case II (green)}: $\xi = \xi_c $. \textbf{The discriminant is equal to zero}. This region represents: critical threshold;
\item \textbf{Case III (red)}: $\xi > \xi_c $. \textbf{The discriminant is negative}. This region represents: anomalous transparency.
\end{itemize}

\begin{figure}[h!]
\label{fig1}
\centering
\includegraphics[scale=0.5]{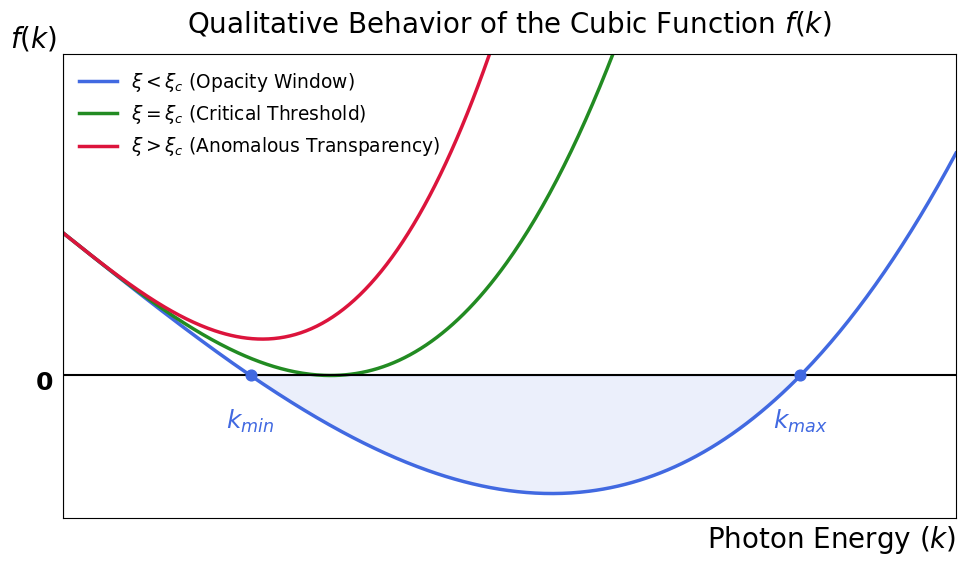}
\caption{Qualitative behavior of the cubic function $f(k)$ in relation to the factors $\xi$ and $\xi_c$. The blue curve, with two real and positive roots, represents the phenomenon called opacity window. The green curve touches the axis at a critical threshold, while the red curve, which has no real and positive solution, represents anomalous transparency.}
\end{figure}
In the case I $(\xi < \xi_c)$, represented by the blue curve, we have an opacity window, that is, where pair production occurs through the Breit-Wheeler effect. In this case, we have two real and positive roots, $k_{min}$ and $k_{max}$. These values define the energy range of the window. Solving the cubic threshold equation in the limit where $\xi \ll 1$, for $k>0$, we have that the minimum threshold is given by
\begin{equation}\label{min}
k_{min} \approx \frac{m_e^2}{\epsilon_b}\left(1 + \frac{\xi m_e^4}{4 \epsilon_b^3}\right),
\end{equation}
since the term $(\xi k^3)$ acts as a first-order perturbation in low-energy regimes. However, for larger values of $k$, the rest mass $(4m_e^4)$ becomes negligible compared to the dynamic terms, giving us
\begin{equation}
k_{max} \approx \sqrt{\frac{4 \epsilon_b}{\xi}}.
\end{equation}
This interval is shaded in blue in Fig. 1. Note that both $k_{min}$ and $k_{max}$ depend on $\xi$.

In case II, represented by the green line, the two roots of case I meet at a single point forming a critical threshold, which is simply the boundary between the allowed and disallowed energy window. FOr last, the case III, given in the graph by the red curve, is called anomalous transparency, where photons pass through without interacting with the background lights and propagate freely.

It is important to emphasize that many works investigating LIV-induced threshold anomalies adopt a generic and simplified MDR model~\cite{matingli1, xiao1, Shao1}, with an arbitrary $\xi$, such as:
\begin{equation}
\omega^2 = k^2 - \xi k^{n},
\end{equation}
where the positive LIV parameter $\xi$ represents the subluminal case, the negative represents the superluminal case, and when it is zero the dispersion relation returns to that of special relativity.

In this research, the development of the cubic threshold function, the determination of the critical point, and the analysis of the discriminants were based entirely on the loop quantum gravity MDR given by Alfaro et al.~\cite{alfaro2002-1}, where the parameter $\xi_{LQG}=-2 \ell_p \theta_7$ is not arbitrary, but fixed by the theory.

Since we have already performed a qualitative analysis of the discriminants and the relationships between $\xi$ and $\xi_c$, we will now proceed with a quantitative analysis of these relationships. Therefore, we will begin the Results and Discussion Section.

\section{Results and Discussion} 
With the theoretical foundation developed, we begin the Results and Discussion Section. Using an EBL model, we will define the energy values for EBL and then for CMB. This is necessary to numerically calculate the critical points of these cosmic backgrounds. Subsequently, we will calculate the Lorentz violation factor of the LQG, given in $\eqref{lq}$, for a quantitative analysis with the critical points of the background lights. This will allow the construction of a table of values, comparing the critical thresholds of special relativity with the adopted LQG model, enabling the interpretation of regions of anomalous opacity and transparency, which will also be visually demonstrated through the geometry of a graph constructed from this table of values.

This section continues with an analysis of the survival percentage of the 1.42 PeV photon detected by LHAASO, connecting our results with an observation in UHE astrophysics. Finally, we will address the gamma-ray horizon from the perspective of special relativity and the LQG.

\subsection{Numerical Calculation of Critical Points}
We will begin the numerical calculation of the critical points of the background radiations. However, to obtain these thresholds, we first need to determine the energy of the EBL and the CMB.

We consider that the best model for analyzing the EBL in the case of pair production is the work of Dominguez~\cite{dominguez2011}, where he measures the spectrum of this background light approximately between:
\begin{equation}
\lambda \approx 0.1 \, \mathrm{\mu m} - 1000 \, \mathrm{\mu m}.
\end{equation}
Within this wavelength range, there are two energy peaks that we will calculate. The first is called the Cosmic Optical Background (COB), which encompasses the optical and near-infrared. This spectrum originates from the direct light of stars accumulated throughout the formation history of all galaxies. It begins in the optical range, but due to the expansion of the Universe, it extends to the near-infrared. The second is called Cosmic Infrared Background (CIB), which encompasses the far-infrared, as it is formed by heated interstellar dust and emits thermal radiation at very low energies. To calculate the energy of these background lights, we will use the relation:
\begin{equation}
\epsilon_{peak} \approx \frac{1.24}{\lambda}.
\end{equation}
The wavelength $\lambda$ must be given in $\mu m$. Thus, the COB peak occurs when
\begin{equation}
\lambda_{COB} \sim 1.1 \, \mathrm{\mu m}.
\end{equation}
That way,
\begin{equation}\label{co}
\epsilon_{COB} \approx \frac{1.24}{1.1} \approx 1.13 \,\mathrm{eV}.
\end{equation}
Now, for the CIB peak, we have
\begin{equation}
\lambda_{CIB} \sim 100 \, \mathrm{\mu m},
\end{equation}
that provides us
\begin{equation}\label{ci}
\epsilon_{CIB} = \frac{1.24}{100} \approx 0.0124 \,\mathrm{eV}.
\end{equation}

To calculate the CMB energy, the best strategy in threshold anomaly studies is to use the frequency based on temperature data from the Planck mission~\cite{planck2018}, and not the wavelength. The peak frequency indicates a higher photon density. Thus, we have
\begin{equation}
\nu \sim 160.23 \, \mathrm{GHz},
\end{equation}
which provides us with the following energy from the CMB:
\begin{equation}\label{cmb}
\epsilon_{CMB} \approx 6.6 \times 10^{-4} \, \mathrm{eV}.
\end{equation}
Since the kinematics of the Breit-Wheeler effect is a stochastic process, in threshold anomalies induced by LIV, this higher photon density at the frequency peak produces a greater possibility of shifting the lower threshold of special relativity.

Now, we will begin the numerical calculation of the critical points, which depend exclusively on the energies of the cosmic backgrounds.

Based on the background energy $\eqref{co}$, we will start with COB:
\begin{equation}
\xi_c^{COB} \approx 1.25 \times 10^{-23} \, \mathrm{eV^{-1}}.
\end{equation}
Through the energy given in $\eqref{ci}$, the CIB is estimated at
\begin{equation}
\xi_c^{CIB} \approx 1.66 \times 10^{-29} \, \mathrm{eV^{-1}}.
\end{equation}
Finally, through the background energy $\eqref{cmb}$, the critical point of the CMB is approximately
\begin{equation}
\xi_c^{CMB} \approx 2.50 \times 10^{-33} \, \mathrm{eV^{-1}}.
\end{equation}

Given that all critical points of the cosmic backgrounds are numerically established, the next step is to calculate the $\xi_{LQG}$ of $\eqref{lq}$, in order to compare it directly with the established critical thresholds.

\subsection{Quantitative Analysis of Discriminants}
To compare the effective Lorentz violation factor of the LQG with the critical points, it is necessary to start with the numerical calculation of $\xi_{LQG}$, established in $\eqref{lq}$. Thus,
\begin{equation}
\frac{1}{\theta_7 \ell_p}=-3.6 \times 10^{17} \, \mathrm{GeV} = -3.6 \times 10^{26} \, \mathrm{eV}.
\end{equation}
Considering $|\theta_7|=33.9$ a good result in the order $\mathcal{O}(10)$, we finally obtain
\begin{equation}
\xi^{LQG} \approx 5, 50 \times 10^{-27} \, \mathrm{eV^{-1}}.
\end{equation}
It is important to emphasize that the first time $\theta_7$ was deduced in the electromagnetic sector of the LQG was in Ref.~\cite{alfaro2002-1} and phenomenologically restricted, possibly, to \cite{Li22}.

Now, we will present Tab. 1 with the critical points of the COB, CIB, and CMB cosmic backgrounds, in order to relate them to the parameter $\xi_{LQG}$ calculated above. Furthermore, it is important to compare the final status of these relationships with the values obtained for the critical thresholds of special relativity.\\
\begin{table}[h!]
\centering
\caption{Parâmetros de massa efetiva \(m_{f}\) e \(2m_{f}\) para diferentes radiações cósmicas de fundo em $E_1 = 1,42$ PeV.}
\label{tab:final_results_1.13}
\begin{tabular}{lccc}
\hline \hline
\textbf{Parameters} & \textbf{COB} & \textbf{CIB} & \textbf{CMB} \\ \hline
Peak $\lambda$ [$\mu$m] & $1.1$ & $100$ & $1060$ \\
Energy $\epsilon$ [eV] & $1.13$ & $0.0124$ & $6.6 \times 10^{-4}$ \\
Critical $\xi_c$ [$\text{eV}^{-1}$] & $1.25 \times 10^{-23}$ & $1.66 \times 10^{-29}$ & $2.50 \times 10^{-33}$ \\
$\xi_{LQG} < \xi_c$? & \textbf{Yes} & \textbf{No} & \textbf{No} \\
SR $k_{min}$ [eV] & $2.31 \times 10^{11}$ & $2.11 \times 10^{13}$ & $3.96 \times 10^{14}$ \\
LQG Window [$k_{min}, k_{max}$] [eV] & $[2.31, 287] \times 10^{11}$ & --- & --- \\
\textbf{Final Status} & \textbf{Opacity Window} & \textbf{Transparent} & \textbf{Transparent} \\ \hline \hline
\end{tabular}
\end{table}\\

The values established in Table 1 are indispensable for explaining phenomena such as anomalous transparency and opacity window generated by the effects of loop quantum gravity. The critical thresholds and energies mentioned are of fundamental importance for the analysis of the graphical behavior in Fig. 2.\\
\begin{figure}[htp!]
\centering
\includegraphics[scale=0.46]{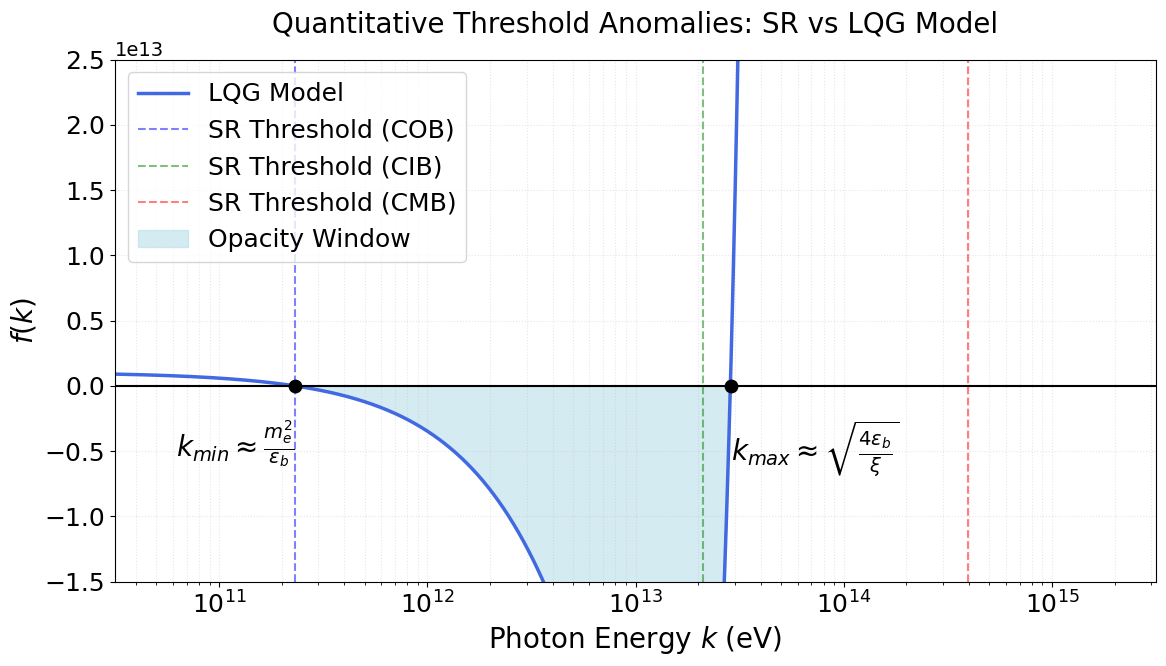}
\caption{Quantitative behavior of the function $f(k)$ in relation to the parameters $\xi_{LQG}$ and $\xi_c$ of the cosmic backgrounds, comparing standard relativity with the   adopted LQG model. The blue curve represents the opacity window generated by the LQG with COB radiation, which opens at $\sim 230$ GeV and closes at $\sim 29$ TeV, the region of which is shaded at $f(k) \leq 0$. The dashed lines in blue, green, and red are the critical points of COB, CIB, and CMB radiation, respectively, for the case of standard relativity. The x-axis is represented on a logarithmic scale for better visualization of the photon's energy evolution.}
\end{figure}\\

Through Fig. 2, we have the geometric behavior of the values mentioned in Tab. 1, which will allow a quantitative analysis of the discriminants of the cubic threshold equation, using the LIV parameter of quantum gravity $(\xi_{LQG} = 5.50 \times 10^{-27} \,\mathrm{eV^{-1}})$. In addition, we trace the threshold energies for the CIB, COB, and CMB cases in SR, clearly showing the difference in the adopted LQG model. This will allow us to make an important observational connection. First, let us establish a scaling factor
\begin{equation}
\eta=\frac{\xi_{LQG}}{\xi_c},
\end{equation}
that allows us to measure how close or far one parameter is from another.

Below is the relationship between $\xi_{LQG}$ and the critical points of the cosmic backgrounds:
\begin{itemize}
\item \textbf{COB: opacity window $(\xi_{LQG} < \xi_c)$}. \textbf{Positive discriminant}: The value of $\eta$ for this relationship is on the order of $10^{-4}$. This means that the effective Lorentz invariance violation factor of the LQG is sufficiently smaller than the critical threshold of the COB peak, allowing the opening of an opacity window, represented in the graph by the shaded region where $f(k) \leq 0$. The negative value of the cubic threshold function can be explained by the relation $4k\epsilon_b - \xi k^3 \geq 4m_e^2$. This tells us that, for an opacity window to exist, the linear term of the collision energy between the gamma photon and the COB radiation photon, given by $(4k\epsilon_b)$, is subtracted by the LQG penalty $(\xi k^3)$, due to the kinematic cost generated by the spacetime discretization for very high-energy photons. The sum of these terms must be greater than or equal to the rest mass $(4m_e^2)$. However, as momentum $k$ grows cubically, this kinematic cost becomes increasingly larger, since VHE photons interact more strongly with the granularity of spacetime. This situation inevitably forces the function to return to the region $f(k)>0$, where the pair production window disappears and photons begin to propagate freely.

As we can observe in the table and in the geometry of the graph itself, the opacity window has a $k_{min}$ and a $k_{max}$. It is now important to establish the width of this window:
\begin{equation}
\Delta k = k_{max} - k_{min} \approx 28.5 \, \mathrm{TeV}.
\end{equation}
In this case, we do not use the $k_{min}$ from the $\eqref{min}$ relation. The value from special relativity, given in $\eqref{sr}$, was sufficient, since the LIV correction $(\xi 	 m_e^4/4\epsilon_b^3)$ the order of $10^{-5}$. This implies that the effective Lorentz invariance violation factor of the LQG slightly shifts the $k_{min}$ of special relativity. However, we consider this difference negligible. The factor that includes new physics and is directly generated by the $\xi$ of the LQG is $k_{max} \approx 28,7 \, \mathrm{TeV}$, which closes the opacity window.

\item \textbf{CIB and CMB: anomalous transparency $(\xi_{LQG} > \xi_c)$}. \textbf{Negative discriminant}: the value of $\eta$ is on the order of $10^2$ and $10^6$ for the CIB peak and for the CMB, respectively. This means that $\xi_{LQG}$ is much larger than the critical points of the CIB and, especially, the CMB, resulting in an anomalous transparency state for gamma photons, which do not interact with these cosmic backgrounds and reach the detectors freely. In the cubic threshold equation, this means that $4k\epsilon_b - \xi k^3 < 4m_e^2$, that is, the kinematic cost $(\xi k^3)$ of the LQG is high enough to prevent pair production. Thus, the Breit-Wheleer effect is kinematically forbidden, resulting in a transparent Universe for VHE and UHE photons.
\end{itemize}

This quantitative analysis is consistent with classic and contemporary references in this broad area of research. Gould and Schréder~\cite{Gould1}, for example, in 1967, were pioneers in investigating the opacity of the Universe to high-energy photons from cosmological sources. Kluzniak~\cite{kluzniak}, in 1999, proposed to explain that modifications in Lorentz invariance, arising, for example, from quantum gravity theories, could establish an upper limit for this opacity of the Universe, while photons with energies above this threshold could propagate freely.

These works laid the groundwork for a more thorough investigation of these phenomena, such as the research by Manttingly et al.~\cite{matingli2} and Jacobson et al.~\cite{jacob1} in 2003, who rigorously and systematically investigated the limits of Lorentz invariance violation, showing that theories predicting LIV, such as various approaches to quantum gravity, must exist at much higher energies than previously thought. In a subsequent work, in 2009, Liberati and Manttingly~\cite{macione1} conducted several studies on quantum gravity phenomenology, investigating how to incorporate Lorentz violation breaks through effective field theory (EFT), rigorously analyzing the feasibility of this approach. These references are highly relevant to this work, since we use an effective quantum gravity model in the semiclassical regime, a hypothesis that predicts a LIV parameter that induces threshold anomalies in the propagation of VHE and UHE photons.

With the well-founded quantitative analysis of the discriminants, we can proceed to an important observational connection.

\subsection{Observational Connections}
We have previously discussed that the effective Lorentz violation factor of the LQG $(\xi_{LQG} = 5.50 \times 10^{-27} \, \mathrm{eV^{-1}})$ imposes an opacity window with a maximum limit for photon attenuation by the COB peak, approximately given by $\sim 29$ TeV $(\Delta >0)$. Furthermore, the CIB peak and the CMB fall into an anomalous transparency regime $(\Delta<0)$, making the Universe transparent for PeV-scale photons to propagate freely.

The most emblematic case regarding the effects of the LQG and the propagation of ultra high energy gamma photons is the detection of the 1.42 PeV photon from the LHAASO J2032+410 source, within the Milky Way, in the region Cygnus Cocoon~\cite{LHAASO2021}. This energy value challenges the opacity of the Universe which, according to special relativity, imposes a critical threshold of $\sim 0.4$ PeV for CMB background photons. That is, in SR, above 0.4 PeV, the Universe is opaque to UHE photons. When LHAASO detects a 1.42 PeV photon, the anomalous transparency imposed by the LQG factor $(\xi k^3)$ is a possible explanation for this phenomenon. Since $\xi_{LQG}$ is much larger than $\xi_c$ of the CMB (about $10^6$ orders of magnitude), it warps spacetime effectively enough to prevent pair formation through the kinematics of the Breit-Wheeler effect. H. Li and Ma, in Ref~\cite{Li4}, meticulously analyze threshold anomalies and investigate the influence of LIV on photons that should be attenuated by cosmic backgrounds but reach the detectors. In Ref~\cite{Li3}, C. Li and Ma use LHAASO data to investigate how the detected UHE photons might signal new physics beyond the Standard Model.

To better understand this, consider the expression for the mean free path of the photon given by
\begin{equation}
\lambda=\frac{1}{n_{bg}.\sigma},
\end{equation}
where $n_{bg}$ is the density of the background photons and $\sigma$ is the cross-section for pair production. Thus, if the Universe is transparent to photons above $\sim 29$ TeV, according to the LQG, the effective cross-section is $\sigma \approx 0$. In this way, the mean free path diverges $(\lambda \rightarrow \infty)$. This means that even if the 1.42 PeV photon detected by LHAASO is within the cosmic gamma-ray horizon of standard relativity, it will not be attenuated by the CMB because, according to the LQG, there is an indefinite extension of this horizon.

To understand Tab. 2, constructed using data from Ling et al~\cite{yling2022} on the LHAASO J2032+4102 source, it is important to define the photon survival factor:
\begin{equation}
P_{\gamma \rightarrow \gamma} = e^{-\tau_{\gamma}(E_0,z_s)}. 
\end{equation}
This formalism, using the mean free path of the photon and the optical depth, was also used by Ruffini et al.~\cite{ruffini1} in 2016, who rigorously calculated to investigate the cosmic opacity of the Universe for UHE photons according to the special relativity model.
\begin{table}[h!]
\centering
\caption{Survival Probability: SR and LQG Model ($\xi \approx 5.50 \times 10^{-27} \, \text{eV}^{-1}$).}
\label{tab:vertical_transparency}
\begin{tabular}{lcc}
\hline \hline
\textbf{Parameters} & \textbf{PeV Regime (CMB)} \\ \hline
Cosmic Background & CMB \\
Source / Event & LHAASO J2032+4102 \\
Photon Energy ($k$) & $1.42$ PeV \\
Source Distance ($d$) & 1.4 kpc \\
Optical Depth ($\tau$) & 0.18 \\  
Mean free path ($\lambda$) & 7.68 kpc \\ 
SR Survival ($P_{SR}$) & $\approx 83\%$ \\ \hline
LQG Mean free path ($\lambda_{LQG}$) & $\infty$ \\
LQG Survival ($P_{LQG}$) & $\approx \mathbf{100\%}$ \\ 
LQG Phenomenon & (Transparency) \\ 
\hline \hline
\end{tabular}
\end{table}

Analyzing Tab. 2, the 1.42 TeV photon, emitted by Cygnus Cocoon, has an attenuation probability of approximately 17$\%$. However, the effects of the $(\xi k^3)$ term are strong enough to prevent the production of pairs between the CMB and photons in the PeV range, so that $\textbf{\textit{P}} \approx 100\%$.

Within the phenomenology of loop quantum gravity, the spacetime distortion (imposed by $\xi_{LQG}$) is strong enough for anomalous transparency without limit, starting from $\sim 29 \ \rm{TeV}$, including for extragalactic photons. This favors the LQG model used to explain the ultra high energy photons detected by LHAASO.

Although we focus on the record-breaking photon on the PeV scale, within the anomalous transparency phenomenon revealed by our LQG model, it is also important to discuss the detection of photons on the very high energy scale, as this reignites the discussion about the "gamma-ray crisis", a term coined by Protheroe and Meyer.~\cite{protery} in 2000 to refer to the detection of TeV photons from the active galaxy Markarian 501, which should be attenuated by the infrared background radiation. Kifune~\cite{kifune}, in 1999, was one of the pioneers in stating that threshold anomalies could be induced by LIV, even discussing the extent of the gamma-ray horizon. Recently, in 2023, H. Li and Ma~\cite{Li5} investigated the detection of multiple TeV from the GRB 221009A source to explain the anomalous transparency for TeV photons using dispersion relations modified by LIV factors. These studies corroborate the phenomena investigated in this work.

We must remember that the GRB 221009A event established a strict upper limit for the linear order of the Lorentz violation factor~\cite{LHAASO_PRL_2024}, estimated at $\xi_{LHAASO} < 8.20 \times 10^{-30} \, \mathrm{eV^{-1}}$. Since the $\xi_{LQG} \approx 5.50 \times 10^{-27} \, \mathrm{eV^{-1}}$ of Alfaro et al. ~\cite{alfaro2002-1} is much larger than the $\xi_{lim}$ of LHAASO, we have an apparent phenomenological tension between these two parameters. However, the upper threshold for the LIV of LHAASO is constrained based on the propagation speed of photons in vacuum as a function of energy, in other words, the time of flight since the GRB emission. Our study, using a LQG model, is based on pair production (Breit-Wheeler effect) through the attenuation of gamma photons by EBL and CMB background radiations. Therefore, the parameter $\xi_{LQG}$ remains necessary to explain the anomalous transparency for the record-breaking 1.42 PeV photon, which would be strongly suppressed by the CMB according to special relativity. This shows that different physical processes can be sensitive to different Lorentz violation parameters.
 
We established a comparison between the SR model and the LQG model adopted to perform the necessary observational connection of a recent detection by the LHASSO detector.

\section{Conclusion}
In this work, we investigate the problem of cosmic photons threshold anomalies induced by the Lorentz invariance violation. Starting from the electromagnetic sector of the LQG, through an effective hamiltonian in the semiclassical regime, we arrive at modified Maxwell equations. Through the kinematics of the Breit-Wheeler effect, we obtained a dispersion relation with a Lorentz invariance violation factor $\xi_{LQG}$. Through this MDR, we arrived at a cubic threshold equation whose discriminant analysis allowed us to find the critical thresholds of cosmic backgrounds (COB, CIB, and CMB), and through the relationship between these parameters, explain phenomena such as opacity window and anomalous transparency.

The highlight of this research lies in the discovery of a strong anomalous transparency for photons on the PeV scale, which should be rigorously suppressed by the cosmic microwave background radiation, since the LIV factor of the LQG is about $10^6$ orders of magnitude greater than the critical threshold of the CMB, contradicting standard relativity, which imposes a limit of $\sim 0.4$ PeV for the free propagation of photons. Clearly, this provides a robust explanation for the detection of the 1.42 PeV photon from the Cygnus Cocoon region.

Another great advantage for our LQG model was the indefinite extent of the gamma-ray horizon, since the opacity window closes at $\sim 29$ TeV, leaving the Universe transparent enough for the detection of UHE photons. As the mean free path limit diverges, not only can galactic photons be detected, such as the 1.42 PeV photon that traveled only 1.4 kpc, but also ultra high energy extragalactic photons traveling much greater distances.

This research demonstrates that loop quantum gravity provides a consistent explanation for threshold anomalies caused by the Lorentz invariance violation and for the phenomenon of anomalous transparency for UHE photons. As a result, we were able to establish an important observational connection with gamma astronomy. We hope that future gamma-ray detectors, such as the CTA, can corroborate the specifications described by our LQG model and confirm signatures of quantum gravity in spacetime.

Our goal is to use this same methodology, which has proven consistent, in other areas of the Standard Model of Particle Physics. We will continue this work by applying the LQG cubic threshold equation to ultra high energy protons to obtain new interpretations of the GZK limit. We also intend to evaluate changes in oscillation probabilities and neutrino decay, providing a unified framework for how our quantum gravity model affects the dynamics of other elementary particles.

\end{document}